\documentclass[aps,prl,twocolumn,showpacs,superscriptaddress,groupedaddress]{revtex4}  
\usepackage{graphicx}  
\usepackage{dcolumn}   
\usepackage{bm}        
\usepackage{amssymb}   

\begin{document}

\widetext
\leftline{Version 01 as of \today}
\author{Marcelo Schiffer\\ Physics Department\\Ariel University  ,Israel}
\leftline{To be submitted to PRD}
\email{schiffer@ariel.ac.il}

\title{ Does the Cosmological Expansion Change  Local Dynamics?}
\date{\today}

\begin{abstract}
It is well a known fact that the Newtonian description of dynamics within Galaxies for its known  matter  content is in disagreement with the observations as the  acceleration approaches   $a_0 \approx 1.2 \times 10^{-10}m/s^2$ (slighter larger for clusters). Both the Dark Matter scenario and Modified Gravity  Theories (MGT)  fails to explain the existence of such an acceleration  scale. Motivated  by the closeness of this acceleration scale  and $c H_0\approx  10^{-9} h$ $ m/s^2$,  we analyse whether this coincidence might have a Cosmological origin for scalar-tensor  and  spinor-tensor  theories,  performing  detailed calculations for perturbations  that represent the local matter distribution on the top of the cosmological background.  Then, we solve the field equations for these perturbations in a power series in the present value of the Hubble constant. As we shall see, for both theories the power expansion contains only even powers in the Hubble constant, a fact that renders the  cosmological expansion irrelevant for the local dynamics.  At last, we show what a difference a theory predicting linear terms in H  makes in the local dynamics. 
\end{abstract}

\maketitle


\section*{Dark Matter or Modified Gravity}

The discrepancy between the Newtonian prediction that  orbital velocities within spiral Galaxies  fall off  as $v \sim (MG/r)^{0.5}$ away from the bulk of the galactic mass distribution  and observations that reveal  that  in every spiral galaxy the velocity distribution  reaches a plateau   as the accelerations approach the value $a_0\approx 1.2 \times 10^{-10} m/s^2$ \cite{milgroma0} led to two diametrically distinct approaches to the conundrum: (i) the  Dark Matter Scenario \cite{dm} where putative non-barionic dark matter with a spherical distribution involving  the disk galaxy provides the needed mass deficit to  conform to the observed flat rotation curves and still adhere to the Newtonian paradigm -- in this case,  the Newtonian potential has a logarithmic dependence on $r$ which is what is needed to provide the flat rotation curves; (ii) Mond Scenario \cite{mond1} , \cite{mond2} in which  the relation between the acceleration and Newtonian gravitational potential is given by
\begin{equation}
\vec{\nabla}\Phi_N =-\mu(a/a_0) \vec{a}
\end{equation}
where $\mu(x)$ is a function  such that $\mu(x)\rightarrow 1$ as $x>>1$ to recover the Newtonian limit and  $\mu(x)\rightarrow x$ as $x<<1$ to reproduce the flat rotation curves of galaxies. One of the immediate consequences of this approach is the automatic reproduction of the Tully-Fisher Law that states that the galaxy luminosity of the galaxy scales  as $ L \sim v^4$ , where $v$ is the orbital  velocity away from the mass distribution, provided that Luminosity tracks the Mass. The defenders of  Mond  claim that in order to the dark matter paradigm to conform to the Tully's -Fisher law , a very  precise (and quite unreasonable ) fine-tuning between the hallo distribution and the observed mass distribution in the galactic disk is required \cite{bekenstein}.

The MOND paradigm evolved into a relativistic equation TeVeS \cite{teves} involving the metric, a scalar and a vector field phrased in terms of a Lagrangian principle. The theory is very successful in reproducing  the rotation curves in spiral Galaxies but is at odds with observed background radiation anisotropies  \cite{microwave}. Furthemore it is in blatant  disagreement with weak lensing observations. The latter is made particularly transparent by the Bullet Cluster lensing observations \cite{bullet1},\cite{bullet2}. 

While the dark matter paradigm cannot explain the existence of the transition acceleration scale $a_0$, in TeVeS it enters as a God-Given parameter in the Lagrangian. Neither one of these possibilities  is theoretically  acceptable.  Intriguingly, $a_0$ comes very close to $c H_0 \approx h 10^{-9} m/s^2$ and  raises the question whether  the change on the dynamical behaviour has a cosmological origin. This avenue was exploited to some degree in the past \cite{nor} ,\cite{feinstein}. 

According to Birkhoff's theorem, in pure Einstein's theory the gravitational field of a spherical symmetric mass configuration is determined by the mass within a sphere of the radius of the observed point alone. Therefore we do not expect the Universe  to  play any role in the local dynamics. A gauge vector field is likewise of no avail;  by Gauss' theorem it also depends upon the internal configuration. Thus if the Cosmological expansion  is to "leak" into the Galactic dynamics,   scalar, spinors or non-gauge vector fields must be called for. 

In this paper we deal with a Brans-Dicke theory and  carefully write down the field equations for linearised perturbations on the top of the cosmological background. In the next section we shall write down these equations in terms  of one scalar field and 3D scalar,  vector and tensor fields. These equations are corrections of the dynamical equations and contain correction terms in powers of $H_0$ . The exact field equations  are then solved perturbatively in powers of $H_0$. The gravitational potential  contains only even powers of $H_0$ and we expand it up to $H_0^4$. It turns out that all corrections are way too small to play any role in the local dynamics.  Then, in the following section we study a massless spinor field and show that also in this case are  no  linear corrections  in $H_0$. Since there is no a priori reason for the absence of odd powers in the Hubble constant, we discuss the prospects  of a linear term in $H_0$ and show that it  brings about noticeable changes the local dynamics .
 
\section{Brans-Dicke Theory}
Brans-Dicke theory   is defined  by the equations of motion 
\begin{equation}
\Box\phi = \frac{8\pi}{3+2\omega}T^M
\label{trace}
\end{equation}
and
\begin{equation}
G_{ab} =  8\pi \left( \frac{T^M_{ab}}{\phi}+ T^\phi_{ab}\right) \quad,
\end{equation}
where
\begin{eqnarray}
T^\phi_{ab}&=&\frac{\omega}{8 \pi\phi^2}
(\nabla_a\phi\nabla_b\phi-\frac{1}{2}g_{ab}\nabla_c\phi\nabla^c\phi)\\
&+&\frac{1}{8\pi \phi}(\nabla_a\nabla_b\phi-g_{ab}\Box\phi)
\end{eqnarray}
and the matter and vacuum energy distributions are represented by
\begin{equation}
T^M_{ab}=(p+\rho)V_a V_b+pg_{ab}
\end{equation}
where  $p_M=0$ and $p_\Lambda =-\rho_\Lambda$, for the present state of the Universe.  Consequently $T^M=-(\rho_M+4\rho_\Lambda)$. For future reference, we recall that
\begin{eqnarray}
\Box \phi=-\ddot{\phi}-3\frac{\dot{a}}{a}\dot{\phi} \quad .
\label{dalambert}
\end{eqnarray}

We wish to construct the field perturbations on the top of a cosmological background for the Brans Dicke Theory;  they  represent the local  matter distribution.  First things first,  we start by solving  the equations for the background fields. In the absence of any dimensional parameter we assume that for a short time interval (the  observation time ) 
\begin{equation}
\frac{\dot{a}}{a}=H \rightarrow \frac{\dot{\phi}}{\phi}=\eta H
\end{equation}
for some dimensionless   $\eta \sim \mathcal{O}(1)$. Then, with this parametrization   
\begin{equation}
T^\phi_{00}=\frac{\eta H^2}{16 \pi} (\omega \eta -6)
\end{equation}
and
\begin{equation}
T^\phi_{\alpha\beta}=\frac{\eta H^2 }{8\pi}\left( \frac{ \omega \eta}{2}+\frac{\dot{H}}{H^2} +2+\eta\right)a^2 \delta_{\alpha\beta} \quad .
\end{equation}
We identify the  energy density and the pressure exerted by the field as
\begin{eqnarray}
\rho_\phi&=&\frac{\eta H^2 \phi }{16 \pi} (\omega \eta -6) \\
p_\phi&=& \frac{\eta H^2 \phi }{8\pi}\left( \frac{ \omega \eta}{2}+\frac{\dot{H}}{H^2} +2+\eta\right) 
\end{eqnarray}

Defining as usual $\rho_c=3H^2 \phi /8 \pi$ and $\Omega_X =\rho_X /\rho_c$,  from Friedmann's equations
\begin{equation}
\Omega_M + \Omega_\Lambda + \frac{\omega \eta^2}{6}-\eta=0
\label{friedmann}
\end{equation}
and
\begin{equation}
\frac{\dot{H}}{H^2}=\frac{3(\Omega_\Lambda-1) -\omega\eta^2/2}{2+\eta} -\eta
\end{equation}
The field  equation for the Brans-Dicke field yields
\begin{equation}
 \frac{\dot{H}}{H^2} =3\frac{\Omega_M +4 \Omega_\Lambda}{(2\omega+3)\eta} -3 -\eta \quad .
\end{equation}

Aiming  solving the perturbed equations, we display Einstein's equations in a more convenient  form
\begin{equation}
R_{ab} =  8\pi \left( \frac{S^M_{ab}}{\phi}+ S^\phi_{ab}\right)
\label{einstein}
\end{equation}
where
\begin{equation}
S^M_{ab}=(p+\rho)V_a V_b+\frac{\rho-p}{2} g_{ab}
\label{smatter}
\end{equation}
and
\begin{equation}
S^\phi_{ab}=\frac{\omega}{8 \pi\phi^2}
(\nabla_a\phi\nabla_b\phi)
+\frac{1}{8\pi \phi}(\nabla_a\nabla_b\phi+\frac{1}{2}g_{ab}\Box\phi) \quad .
\label{sfield}
\end{equation}

There are two relevant coordinate systems,  the $r$-frame  ($r^a$ coordinates) locally attached to the local mass distribution  and the  $x$-frame ($x^a$ coordinates) which is the  cosmological comoving frame,   with $r^\alpha =a(t) x^\alpha$. The  r-frame is the physically meaningful   frame for local dynamics but the x-frame  turns out to be much more convenient for performing calculations.  Accordingly, we construct static  local disturbances in the $r$-frame (we are not interested in galactic evolution), make a coordinate transformation to the $x$-frame and perform  calculations, obtaining the perturbed fields.  Then, we transform them back to the $r$-frame. Let $h_{ab}(\vec{r})$ represent the static metric perturbations in the $r$-frame, then the line element is
\begin{equation}
ds^2= \left(g^{(0)}_{ab}+h_{ab}(\vec{r})\right) dr^a dr^b
\end{equation}
where $dr^0=dt$ , $g^{(0)}_{ab}$ is the cosmological smooth background. Under a '$r$' to '$x$' coordinate  transformation the line element perturbation looks 
\begin{widetext}
\begin{equation}
h_{ab}(\vec{r})dr^a dr^b = \left[h_{00} +2H  
h_{0\alpha}r^\alpha  + H^2 h_{\alpha \beta}r^\alpha r^\beta\right]dt^2+2a \left[h_{0\alpha} + H h_{\alpha \beta}  r^\beta \right]dx^\alpha dt+a^2 h_{\alpha\beta} dx^\alpha dx^\beta
\label{tildeh}
\end{equation}
\end{widetext}
where $H=\dot{a}/a$ and we recall that $h_{ab}(\vec{r})=h_{ab}(a \vec{x})$.

Inspecting this form, we  express the perturbed metric in the $x$-frame $\tilde{h}_{ab}$ in the form:
$ \psi, W_\alpha$ and $f_{\alpha\beta}$
\begin{eqnarray}
\tilde{h}_{00}&=&\psi(a\vec{x}) \quad ; \nonumber \\
\quad\quad\tilde{h}_{0\alpha}&=&a W_\alpha(a\vec{x}) ;\nonumber \\
 \tilde{h}_{\alpha\beta}&=&a^2 f_{\alpha\beta}(a \vec{x})
\label{metric}
\end{eqnarray}
where $\psi$, $W_\alpha$ and $f_{\alpha\beta}$  are to be regarded as scalar, vector and tensor fields  of a  flat three dimensional space . It is reasonable to assume that the global space curvature is unimportant on a local scale, thus locally we take $g^{(0)}_{\alpha \beta}=a^2 \delta_{\alpha\beta}$.  Similarly, the perturbation of the scalar field is static in the physical frame ,$\phi +\delta \phi= \phi\left[ 1+   \xi (a x^\alpha)\right]$.

We represent the local mass distribution as a disturbance of the global smooth distribution. In this case, $\delta p$  stands for the pressure and $\delta \rho $  the  mass density of the local matter distribution. Locally  $\delta p=0$ and $\delta \rho=\rho_G $ , the local Galactic mass distribution.  There is still  one missing field $u_a$, the  difference between the velocity of locally static observer in the $r$-frame with respect to a cosmological comoving observer. For a static observer in the local frame 
$x^\alpha = a^{-1} r^\alpha$ with  constant $r^\alpha$. Thus the corresponding  velocity in the $x$-frame is: 
\begin{equation}
V^a_G =\frac{(1;-H\vec{x})}{\sqrt{1-H^2a^2x^2}} \approx (1, -H\vec{x}) \quad .
\end{equation}
Recalling that $V^b$ is the velocity of the cosmological comoving observer
Clearly 
\begin{equation}
u_a =g_{ab} \left (V_G^b-V^b \right)+ \tilde{h}_{ab}V^b
\end{equation}
or
\begin{equation}
u_a =(0,-a H \vec{r})+\tilde{h}_{a0}=\left(\psi,a(-H r^\alpha +W_\alpha)\right) \quad .
\label{u}
\end{equation}

Preparing the ground for calculating the perturbations of the field equations we first evaluate, 
\begin{equation}
\delta\left(\nabla_a\nabla_b\phi\right)=\xi \nabla_a\nabla_b  \phi +\phi \nabla_a\nabla_b  \xi + \nabla_a\xi \nabla_b\phi+\nabla_b\xi \nabla_a\phi -\gamma^c_{ab}\phi_c
\label{abfluctuation}
\end{equation}
where 
\begin{equation}
\gamma^c_{ab}=\delta \Gamma^c_{ab}= \frac{1}{2}\left( \nabla_b \tilde{h}^c_a+\nabla_a \tilde{h}^c_b -\nabla^c \tilde{h}_{ab}\right)
\end{equation}
and consequently 
\begin{equation}
\delta\left(\Box \phi\right)=\xi \Box  \phi +\phi \Box  \xi +2 \nabla_c\xi \nabla^c\phi-\tilde{h}^{cd}\nabla_c\nabla_d\phi -\gamma^c\phi_c
\label{boxvar}
\end{equation}
with $\gamma^c=g^{ab} \gamma^c_{ab}$.  We adopt the Lorentz gauge condition, 
\begin{equation}
\nabla_c \tilde{h}_a^c-\frac{1}{2}\nabla_a \tilde{h}=0 ,
\label{gauge}
\end{equation}
in which case
\begin{equation}
\gamma^c=g^{ab}\gamma^c_{ab}=0 \quad ת
\end{equation}
and simply drop the last term in eq. (\ref{boxvar}). We can express this gauge condition in terms of the effective 3D-fields:
\begin{eqnarray} 
\frac{1}{a}W_{\alpha,\alpha}&=&\frac{1}{2}(\dot{f}+\dot{\psi}) +H (f+3\psi)  \nonumber \\
\frac{1}{a}f_{\alpha\beta,\beta}& =&\dot{W}_\alpha +4H W_\alpha +\frac{1}{2a}(f-\psi)_{,\alpha}
 \label{gaugew}
\end{eqnarray}

The field equations governing the local  scalar field is 
\begin{equation}
\xi \Box \phi +\phi \Box \xi +2 \nabla^a\xi \nabla_a \phi-\tilde{h}^{ab}\nabla_a\nabla_b \phi= \frac{8\pi}{2\omega+3} \delta T
\end{equation}
But

\begin{equation}
\tilde{h}^{cd}\nabla_c\nabla_d \phi=\tilde{h}^{00}\ddot{\phi}-\tilde{h}^{\alpha \beta} \Gamma^0_{\alpha \beta} \dot{\phi}=-\psi\Box \phi-3H\dot{\phi}\psi -H \dot{\phi}f 
\end{equation}
where $f\equiv \sum_\alpha f_{\alpha\alpha}$. Then 
\begin{equation}
(f+\psi) \Box \phi +\phi \Box \xi -2\dot{\phi}\dot{\xi}+H(f+3\psi) \dot{\phi}=\frac{8\pi}{2\omega+3} \delta T
\end{equation}

Clearly  $\delta T=\delta (3 p-\rho)=-\rho_G$ is the local energy already discussed. From eq. (\ref{trace})  
\begin{eqnarray}
& &(f+\psi) \frac{8\pi}{(2\omega+3)\phi}(-\rho+3p)+ \Box \xi -2\frac{\dot{\phi}}{\phi}\dot{\xi} \nonumber  \\ &+& H(f+3\psi) \frac{\dot{\phi}}{\phi}=- \frac{8\pi}{(2\omega+3)\phi} \rho_G \quad .
\label{fieldeqxi}
\end{eqnarray}

We translate back  our equations in terms of r-frame variables. In contrast to  the comoving derivative $\xi_{,\alpha}=  \partial \xi /\partial x^\alpha$  we define the local derivative  $\partial_\alpha \xi  =\partial \xi /\partial {r^\alpha}$. Then
\begin{equation}
[\xi(a x^\alpha)]_{,\alpha}=a \partial_\alpha [\xi(r^\alpha)]
\end{equation}
and as the rule of the thumb we automatically replace everywhere $\partial  /\partial x^\alpha \rightarrow a \partial  /\partial {r^\alpha}$. 
Furthermore
 \begin{equation}
\frac{\partial \xi(a\vec{x})}{\partial t}= H \vec{r}\cdot \vec{\partial} \xi
\label{tder}
\end{equation}
Then
\begin{equation}
\Box \xi(a \vec{x})= (\delta_{\alpha \beta}-H^2 r_\alpha r_\beta)\partial_\alpha\partial_\beta \xi -4 H^2 \vec{r}\cdot\vec{\nabla} \xi
\label{boxi}
\end{equation} 
With the   the replacement 
\begin{equation}
\dot{\phi}/\phi \rightarrow \eta H  \quad;\quad  \phi^{-1}\rightarrow G \quad  \mbox{and}\quad  8\pi \rho_c /\phi  \rightarrow 3H^2 
\label{replacement}
\end{equation}
 the scalar field equation (eq. (\ref{fieldeqxi})) looks in its final form
\begin{widetext}
\begin{equation}
H^2\left[\frac{3}{(2\omega+3)} (\Omega_M+4 \Omega_\Lambda)(\xi+\psi)-(f+3\psi)\eta + 2(\eta+2)   \vec{r}\cdot \vec{\partial} \xi\right]  - (\delta_{\alpha \beta}-H^2 r^\alpha r^\beta)\partial_\alpha\partial_\beta \xi=  \frac{8\pi G}{2\omega+3}\rho_G \quad .
\label{eqxi}
\end{equation}
\end{widetext}

The field equations for the gravitational field are given by the linear perturbations of Einstein's equations:
\begin{equation}
\delta R_{ab} =  8\pi \delta S_{ab}
\label{vareinstein}
\end{equation}
where
\begin{equation}
\delta S_{ab} \equiv   \frac{\delta S^M_{ab}-\xi S^M_{ab}}{\phi}+ \delta S^\phi_{ab} \quad .
\end{equation}

Let me start with the lhs. We borrow from MTW   \cite{MTW}:
\begin{equation}
\delta R_{ab}=\frac{1}{2}\left(-\nabla_a\nabla_b \tilde{h} -\nabla^c\nabla_c \tilde{h}_{ab}+\nabla^c \nabla_a \tilde{h}_{bc}+\nabla^c \nabla_b \tilde{h}_{ac}\right),
\end{equation}
and rewrite the divergence of the gauge condition [eq. (\ref{gauge})] in the form
\begin{equation}
 \nabla_c\nabla_b \tilde{h}^c_a= \frac{1}{2}\nabla_b\nabla_a \tilde{h}+\left[\ \nabla_c\nabla_b-\nabla_b\nabla_c  \right] \tilde{h}^c_a \quad .
\end{equation}
With  the rule for  the commutation of derivates for $(1,1)$ tensors
\begin{equation}
\left[\ \nabla_c\nabla_b-\nabla_b\nabla_c  \right] \tilde{h}^c_{ a} =R_{db}    \tilde{h}^d_{ a}+R_{adcb}    \tilde{h}^{cd}
\end{equation}
it follows that
\begin{equation}
 \nabla_c\nabla_b \tilde{h}^c_a+ \nabla_c\nabla_a \tilde{h}^c_b  = \nabla_b\nabla_a \tilde{h}+R_{db}    \tilde{h}^{da}+R_{da}   \tilde{h}^{db}+2 R^a_{dcb}    \tilde{h}^{cd} ,
\end{equation}
and then
\begin{equation}
\delta R_{ab}=\frac{1}{2}\left( R_{db}    \tilde{h}_a^d+R_{da}   \tilde{h}^d_b+2 R_{adcb}    \tilde{h}^{cd}-\nabla^c\nabla_c \tilde{h}_{ab}\right) .
\end{equation}
This expression is quite general. For a  homogenous and isotropic background  the Weyl tensor vanishes, and  the Riemann tensor is  entirely  described by the Ricci curvature :
\begin{eqnarray}
R_{adcb}&=&\frac{1}{2}\left(g_{ac}R_{db}-g_{ab}R_{cd}-g_{dc}R_{ba}+g_{bd}R_{ca}\right) \\ &+& \frac{1}{6}R\left(g_{ab}g_{cd}-g_{ac}g_{db}\right).
\end{eqnarray}
In that case
\begin{eqnarray}
\delta R_{ab}&=& R_{cb}    \tilde{h}_a^c+R_{ca}   \tilde{h}^c_b-\frac{1}{2}\left(g_{ab}R_{cd}\tilde{h}^{cd}+\tilde{h} R_{ab}\right) \\
&-&\frac{1}{6}(\tilde{h}_{ab}-g_{ab}\tilde{h})R-\frac{1}{2}\nabla_c\nabla^c \tilde{h}_{ab} \quad.
\label{varR}
\end{eqnarray}
Our next step, is to express $\delta R_{ab}$ in terms of the fields $f_{\alpha\beta},W_\alpha$ and $\psi$ according to their definitions [eq.(\ref{metric})]. Furthemore we use the field equations of the unperturbed fields  [eqs. (\ref{einstein}),(\ref{smatter}) and (\ref{sfield})] obtaining 
\begin{widetext}
\begin{eqnarray}
\delta R_{00}&=&-\left[ 8\pi \frac{3\omega p +(\omega+3) \rho }{2\omega+3)\phi} +\omega\frac{\dot{\phi}^2}{\phi^2}-3H \frac{\dot{\phi}} {\phi}\right] \psi
-\left[    
 8\pi\frac{(1+\omega/3) \rho +\omega p}{(2\omega+3)\phi}   
+\frac{\omega}{3}\frac{\dot{\phi}^2}{\phi^2}
-H \frac{\dot{\phi}} {\phi}\right]f-\frac{1}{2} \nabla^c\nabla_c \tilde{h}_{00}
\label{r00}\\
\delta R_{0\alpha}&=&-a\left[8 \pi\frac{(2+\omega/3) \rho+3\omega p}{(2\omega+3)\phi} +\frac{5\omega}{6}\frac{\dot{\phi}^2}{\phi^2}-2H \frac{\dot{\phi}}{\phi}
\right]  W_{\alpha}-\frac{1}{2} \nabla^c\nabla_c \tilde{h}_{0\alpha}
\label{r0alpha}\\
\delta R_{\alpha\beta}&=&a^2 \left[ 8 \pi\frac{(\frac{5\omega}{3}+2)\rho-\omega p}{(2\omega+3)\phi} -2H \frac{\dot{\phi}}{\phi}+\frac{\omega}{6}\frac{\dot{\phi}^2}{\phi^2}\right] f_{\alpha\beta}
-a^2 \delta_{\alpha\beta}\left[ \frac{8 \pi}{(2\omega+3)\phi}  \left[(\omega/3+1)\rho +\omega p) \right]-H \frac{\dot{\phi}}{\phi}+\frac{\omega}{3}\frac{\dot{\phi}^2}{\phi^2}\right]\psi  \nonumber \\
&-&a^2 \delta_{\alpha\beta}\left[ \frac{8 \pi}{3\phi}\rho 
-H \frac{\dot{\phi}}{\phi}+\frac{\omega}{6}\frac{\dot{\phi}^2}{\phi^2}\right]f  -\frac{1}{2} \nabla^c\nabla_c \tilde{h}_{\alpha\beta}  .
\label{ralphabeta}
\end{eqnarray}

Furthermore,
\begin{eqnarray}
 \nabla^c\nabla_c \tilde{h}_{00}&=&a^{-2} \nabla^2  \psi-\ddot{\psi}-3H\dot{\psi}+6H^2 \psi -4H a^{-1} W_{\alpha,\alpha} +2H^2 f
\label{h00}
\\
\nabla^c\nabla_c \tilde{h}_{\alpha0}&=&a\left[a^{-2} \nabla^2 W_\alpha -\ddot{W_\alpha}-3H\dot{W}_{\alpha} +6 H^2 W_\alpha -2H a^{-1}\psi_{,\alpha}-2H a^{-1} f_{\alpha \beta,\beta}\right]
\label{h0alpha}
\\
\nabla^c\nabla_c \tilde{h}_{\alpha\beta}&=&a^2\left[ a^{-2}\nabla^2 f_{\alpha \beta}-\ddot{f_{\alpha\beta}}-3H \dot{f}_{\alpha \beta}+2H^2 f_{\alpha \beta} -2H a^{-1} (W_{\alpha,\beta}+W_{\beta,\alpha}) +2H^2 
\psi \delta_{\alpha\beta} \right] \quad .
 \label{halphabeta}
\end{eqnarray}

The linear variation of eqs. (\ref{smatter}) and (\ref{sfield}) provide  the  source terms of the gravitational field equations: 
\begin{equation}
\delta S^M_{ab}=\rho_M (V_a u_b+V_b u_a)+\frac{2\rho_\Lambda +\rho_M}{2} \tilde{h}_{ab}+\rho_G(V_a V_b+\frac{1}{2} g_{ab})
\label{deltam}
\end{equation}
together with
\begin{equation}
\delta S^\phi_{ab}=\frac{\omega+1}{8 \pi\phi} \left(\nabla_a\xi\nabla_b\phi+\nabla_b\xi\nabla_a\phi\right)
+\frac{1}{8\pi}  \nabla_a\nabla_b\xi-\frac{1}{2\phi(2\omega +3)}\left( g_{ab} \rho_G +(\tilde{h}_{ab}-g_{ab} \xi)(4\rho_\Lambda +\rho_M)\right)-\frac{1}{8\pi} \gamma^0_{ab} \frac{\dot{\phi}}{\phi} \quad .
\label{deltaphi}
\end{equation}
Working out the  components

\begin{eqnarray}
8 \pi\delta S_{00}&=&\frac{\dot{\phi}}{2\phi} \dot{\psi} +
 \ddot{\xi}+2 (\omega+1) \frac{\dot{\phi}}{\phi} \dot{\xi} 
+\frac{8\pi}{\phi}
 \left[\frac{\omega+2}{2\omega+3}\rho_G +
 \frac{(2\omega+1)\rho_\Lambda -(3\omega+5)\rho_M}{2\omega+3} \psi +
\frac{(2\omega+1) \rho_\Lambda -(\omega+2)\rho_M}{2\omega+3}  \xi  \right] 
\label{s00}\\
8\pi \delta S_{\alpha 0} &=&  a \left[\frac{8\pi}{\phi}\frac{(2\omega+1)\rho_\Lambda-(\omega+2)\rho_M}{2\omega+3}-H  \frac{\dot{\phi}}{\phi}   \right] W_\alpha +a\frac{8\pi}{\phi}\rho_M H  r_\alpha +  \left[ (\omega+1) \frac{\dot{\phi}}{\phi} -H\right]\xi_{,\alpha}+ \dot{\xi}_{,\alpha}
+\frac{\dot{\phi}}{2\phi}\psi_{,\alpha} 
\label{s0alpha}\\
8\pi \delta S_{\alpha\beta}&=&a^2\left[\frac{8\pi}{\phi} \frac{(2\omega+1)\rho_\Lambda+(\omega+1)\rho_M}{2\omega+3} -H \frac{\dot{\phi}}{\phi}\right] f_{\alpha\beta}
+a \frac{\dot{\phi}}{2\phi}\left( W_{\alpha,\beta}+W_{\beta,\alpha}\right)\\&-&
a^2 \frac{\dot{\phi}}{2\phi} \dot{f}_{\alpha\beta}+a^2\left[\frac{8\pi}{\phi}
\frac{\omega+1}{2\omega+3}\rho_G -\frac{8\pi}{\phi}\frac{(\omega+1)\rho_M+(2\omega+1) \rho_\Lambda}{2\omega+3}\xi-H\dot{\xi}-H \frac{\dot{\phi}}{\phi} \psi\right]\delta_{\alpha\beta}+ \xi_{,\alpha \beta}
\label{salphabeta}
\end{eqnarray}
\end{widetext}

We shall put all the pieces together ,(\ref{r00})-(\ref{ralphabeta}) with  eqs.  (\ref{h00})-(\ref{halphabeta}) and (\ref{s00})). We use the gauge conditions (eqs.(\ref{gaugew}))  and the replacements (\ref{tder}),  (\ref{replacement}). The 'scalar equation' that arises from the $00$ component is
\begin{widetext}
\begin{eqnarray}
& &H^2 \left(A \psi +B f + C\xi\right) +H^2(3-\frac{1}{2}\eta) \vec{r}\cdot\vec{\partial} \psi+H^2  \vec{r}\cdot \vec{\partial} f -H^2\left(2(\omega+1)\eta +1  \right) {\vec{r}\cdot\vec{\partial} \xi}-H^2
 r_\alpha r_\beta \partial_\alpha \partial_\beta \xi \\
&-&\frac{1}{2}(\delta_{\alpha \beta}- H^2 r_\alpha r_\beta) \partial_\alpha \partial_\beta \psi =\frac{\omega+2}{2\omega+3}8 \pi G  \rho_G
\label{eqpsi}
\end{eqnarray} 
while the vector equation that arises from the $0\alpha$ component is
\begin{equation}
H^2 D W_\alpha -\frac{1}{2}\left(\delta_{\beta \gamma}-H^2 r_\beta r_\gamma \right)      \partial_\beta \partial_\gamma W_\alpha + 
3 H^2 \vec{r}\cdot \vec{\partial} W_\alpha +\frac{1}{2} H \partial_\alpha f +\frac{1-\eta}{2}H  \partial_ \alpha \psi 
-H(\omega+1) \eta \partial_\alpha \xi -H \vec{r}\cdot \vec{\partial} \partial_\alpha \xi= 3  \Omega_M H^3 r_\alpha  .
\label{eqW}
\end{equation}

Last, the tensor equation from the $\alpha\beta$ component can be simplified with the aid of Friedman's equation [ eq.(\ref{friedmann})]
\begin{equation}
-H^2  f_{\alpha\beta}+  H^2 \left( P \psi +Q \xi +\vec{r}\cdot\vec{\partial }\xi \right)   \delta_{\alpha \beta} +
H\left(\frac{2-\eta}{2}\right)  \left(\partial_\alpha W_\beta+\partial _\beta W_\alpha\right) +\left( \frac{\eta}{2}+2 \right)H^2 \vec{r}\cdot\vec{\partial} f_{\alpha \beta}
 -\partial_\alpha \partial_\beta \xi-\frac{1}{2} \left(\delta_{\mu \nu}-H^2 r_\mu r_\nu\right) 
\partial_\mu\partial_\nu f_{\alpha \beta}=\frac{\omega+1}{2\omega+3} 8\pi G \rho_G \delta_{\alpha\beta}
\label{eqf}
\end{equation}
\end{widetext}
where we defined the numerical coefficients coefficients :
\newpage

\newpage
\begin{eqnarray}
A&=&6 \frac{(\omega+1)\Omega_M-2\Omega_\Lambda}{2\omega+3}-\omega\eta^2 +3(\eta+1)\\
B&=&-\frac{(\omega+3)\Omega_M+(3-2\omega)\Omega_\Lambda}{2\omega+3}-\frac{\omega\eta^2}{3}+\eta+1\\
C&=&3 \frac{(\omega+2)\Omega_M -(2\omega+1)\Omega_\Lambda}{2\omega+3} \\
D&=&\frac{2\omega\Omega_M+(2\omega-9) \Omega_\Lambda}{2\omega+3}-5 \frac{\omega \eta^2}{6}+3\eta-1\\
P&=&\frac{(\omega+3)\Omega_M  -(2\omega-3)\Omega_\Lambda}{2\omega+3} -2\eta +\frac{\omega}{3}\eta^2 +1\\
Q&=&3\frac{(\omega+1)\Omega_M  +(2\omega+1)\Omega_\Lambda}{2\omega+3}
\end{eqnarray}

\section*{Solving the equations by Perturbation}

At this stage a remark of caution is in order.
 Albeit the perturbation fields $\psi$, $W_\alpha$ and $f_{\alpha \beta}$ stand for  $\tilde{h}_{\alpha\beta}$ ,  are functions of the local coordinate $\vec{r}$, they are still metric perturbations in the $x$-frame [see eqs. (\ref{tildeh}),(\ref{metric})]: 
\begin{equation}
ds^2=\tilde{g}^{(0)}_{ab}dx^a dx^b + \psi dt^2+2 a W_\alpha dx^\alpha dt +a^2 f_{\alpha\beta} dx^\alpha dx^\beta
\end{equation}
Transforming back to the $r$-frame:
\begin{widetext}
\begin{equation}
ds^2=g^{(0)}_{ab}dr^a dr^b -(\psi +2 H W_\alpha r^\alpha -H^2 f_{\alpha \beta}r^\alpha r^\beta) dt^2 +2(W_\alpha -H f_{\alpha \beta} r^\alpha )dr^\alpha dt +f_{\alpha\beta} dr^\alpha dr^\beta
\end{equation}
\end{widetext}
Clearly
\begin{eqnarray}
h_{00} &=&- \psi -2 H W_\alpha r^\alpha +H^2 f_{\alpha \beta}r^\alpha r^\beta \nonumber \\
h_{0\alpha}&=&W_\alpha -H f_{\alpha \beta} r^\beta \nonumber \\
h_{\alpha \beta}&=&f_{\alpha\beta}
\label{translate}
\end{eqnarray}
We shall consider spherically symmetric configurations alone. In this case
\begin{equation}
W_\alpha =W(r ) \hat{r}_\alpha \quad ; \quad f_{\alpha \beta}=A(r)\delta_{\alpha\beta}+ B(r ) \hat{r}_\alpha \hat{r}_\beta
\end{equation}
where $A, B$ and $W$ are 'scalar fields'. Then
\begin{eqnarray}
\partial^2 W_\alpha &=&\left( \partial^2 W -2\frac{W}{r^2}\right) \hat{r}_\alpha\\
\partial^2 f_{\alpha \beta}&=&\left( \partial^2 A + \frac{2B}{r^2}\right)\delta_{\alpha\beta}+ \left( \partial^2 B  -\frac{6B}{r^2}\right)\hat{r}_\alpha\hat{r}_\beta
\end{eqnarray}
and also
\begin{eqnarray}
\partial_\alpha W_\alpha &=& W'+\frac{2 W}{r} \nonumber \\
\partial_\beta f_{\alpha \beta}&=& \left(A'+B'+\frac{2B}{r}\right) \hat{r}_\alpha
\label{divw}
\end{eqnarray}
Next we introduce these expressions  into the their corresponding  equations  (\ref{eqpsi}) , (\ref{eqpsi})-(\ref{eqf}) and solve them pertubatively  in powers of $H$. The  zeroth order satisfying the gauge conditions is
\begin{eqnarray}
\xi^{(0)}&=&\frac{1}{2\omega+3} \frac{2MG}{r}\\
\psi^{(0)}&=& \frac{\omega+2}{2\omega+3} \frac{4G M}{r}\\
f_{\alpha\beta}^{(0)}&=&\frac{2GM}{r}\delta_{\alpha\beta}+\frac{\omega+1}{2\omega+3}\frac{2 MG}{r} \hat{r}_\alpha \hat{r}_\beta \\
W_\alpha^{(0)}&=&0
\end{eqnarray}

The easiest way of getting $W$ is by substituting  the previous results into the gauge condition (\ref{gaugew}) . From now on we drop numerical coefficients, then
\begin{equation}
\partial_\alpha W^{(1)}_\alpha \sim \frac{ MG}{r}
\end{equation}
and by virtue of (\ref{divw}) it follows that $W \sim MG$ and no r-dependence and then
\begin{equation}
W^{(1)}_\alpha \sim MG \hat{r}^\alpha  .
\end{equation}

To the second order we have 
\begin{widetext}
\begin{equation}
\partial^2 \psi^{(2)} = 2\left(A \psi^{(0)} +B f^{(0)}+ C\xi^{(0)}\right) +
2(3-\frac{1}{2}\eta) \vec{r}\cdot\vec{\partial} \psi^{(0)}+
2\vec{r}\cdot \vec{\partial} f^{(0)}
- 2\left(2(\omega+1)\eta +1  \right) {\vec{r}\cdot\vec{\partial} \xi^{(0)}}
+r_\alpha r_\beta \partial_\alpha \partial_\beta( \psi^{(0)}- 2\xi^{(0)})\nonumber
\end{equation} 
\begin{equation}
\partial^2 f^{(2)}_{\alpha \beta}=2f^{(0)}_{\alpha\beta}+  2 \left( P \psi^{(0)} +Q \xi^{(0)} +\vec{r}\cdot\vec{\partial }\xi^{(0)} \right)   \delta_{\alpha \beta} +
\left(2-\eta \right)  \left(\partial_\alpha W^{(1)}_\beta+\partial _\beta W^{(1)}_\alpha\right) +\left( \eta+4 \right) \vec{r}\cdot\vec{\partial} f^{(0)}_{\alpha \beta}
-2 \partial_\alpha \partial_\beta \xi^{(0)} +  r_\mu r_\nu 
\partial_\mu\partial_\nu f^{(0)}_{\alpha \beta}
\label{2f}
\end{equation}
and 
\begin{equation}
\partial^2 \xi^{(2)}=\frac{3}{(2\omega+3)} (\Omega_M+4 \Omega_\Lambda)(\xi^{(0)}+\psi^{(0)})-\eta (f^{(0)}+3\psi^{(0)})+2(\eta+2)   \vec{r}\cdot \vec{\partial} \xi^{(0)}   +  r^\alpha r^\beta \partial_\alpha\partial_\beta \xi^{(0)}
\label{2xi}
\end{equation}
\end{widetext}
whose solution is
\begin{eqnarray}
\psi^{(2)} &\sim& MG r ;  \\
\xi^{(2)} &\sim& MG r   \nonumber \\ 
 f^{(2)}_{\alpha \beta}  &\sim& M G r( \delta_{\alpha\beta}+\hat{r}_\alpha\hat{r}_\beta) \nonumber
\end{eqnarray}
at higher orders
\begin{widetext}
\begin{eqnarray}
 \partial^2 W^{(3)}&=&2 W^{(1)}_\alpha + r_\beta r_\gamma\partial_{\alpha}\partial_\beta W^{(1)}     + 
6  \vec{r}\cdot \vec{\partial} W^{(1)}_\alpha +  \partial_\alpha f^{(2)} +(1-\eta)  \partial_ \alpha \psi^{(2)} \nonumber \\
&-&2 (\omega+1) \eta \partial_\alpha \xi^{(2)} -2 \vec{r}\cdot \vec{\partial} \partial_\alpha \xi^{(2)}- 6  \Omega_M  r_\alpha 
\end{eqnarray}
\end{widetext}
Acccordingly, 
\begin{equation}
W^{(3)}_\alpha \sim ( MG r^2 - r^3 )\hat{r}_\alpha .
\end{equation}
The fourth order equations for $\psi^{(4)}$ and $\xi^{(4)}$ are identical to (\ref{2f}),(\ref{2xi}) and therefore 
\begin{equation}
\psi^{(4)}\sim MGr^3 \quad; \quad \xi^{(4)}\sim MGr^3 .
\end{equation}
Thus by virtue of  eq. (\ref{translate}), 
\begin{eqnarray}
g_{00}  &\sim& -1 +H^2 r^2 +\frac{\omega+2}{2\omega+3} \frac{4G M}{r}  + H^2 MG r \nonumber \\ & - &H^4 r^4 + H^4 M G r^3+ \dots
\end{eqnarray}

The term $H^2r^2$ arrives from the coordinate transformation   from the $x$ frame to $r$-frame [see eq (\ref{translate})]. Comparison with the Newtonian potential term $GM/r$ tells that it  becomes relevant as $r^3 \sim MG H^{-2}$ or $r \sim 400kpc $ for a typical galaxy. On the same grounds, he correction $H^2 MG r$  becomes relevant only at the Hubble distance  $r \sim  H$  . Notice that there are no linear terms on $H$ that could bring about relevant corrections to the local dynamics.

\section{Spinor Field}
In Brans-Dicke theory the  lowest order  in $H$ corrections of the field equations are quadratic in  the Hubble constant. We  wonder if a spinor field, whose energy momentum tensor contains first derivatives of the spinor field could remedy the problem and yield larger contributions.  Since we agreed upon not to settle the scale of $a_0$ through external given parameters,  we concentrate on a massless particle. All non-zero momentum modes can be swept into the energy momentum tensor of the matter distribution and  the discussion is similar to that of the previous section. Nevertheless,  the zero mode has no  particle content and must be dealt separately.  We think this mode as being a cosmological substrate that is deformed in the presence of a mass distribution and calculate its contribution to the energy-momentum tensor.

In a curved space- time the Dirac equation reads
\begin{equation}
\left[i \gamma^a e_{(a)}^m \left(\frac{ \partial}{\partial x^m} +\frac{1}{4} C_{mbc} \gamma^b \gamma^c \right)  -m \right]\Psi=0
\end{equation}
where $e^{(a)}_m$ $a=1,\dots,4$ are the four tetrads (the index in bracket is a Lorentz index and the other one is the space-time component), 
\begin{equation}
g_{mn}=e^{(a)}_m e^{(b)}_n \eta_{ab} ; 
\end{equation}
Bracketed indexes of the tetrads are raised/lowered with $\eta_{ab}$, unbracketed indexes with the space-time metric $g_{mn}$.
and $\gamma^a$ are the Dirac matrixes 
\begin{equation}
\left\{\gamma^a,\gamma^b\right\}=2\eta^{ab}
\end{equation}
 and the spin connection is defined as
 \begin{equation}
 C_{m(a)(b)}= e_{( a )}^{\quad n} e_{(c ) n;m}
\end{equation}
Furthermore, one defines  the derivative operator 
\begin{equation}
D_m=  \frac{ \partial}{\partial x^m} +\frac{1}{4} C_{mbc} \gamma^b \gamma^c .
\end{equation}
 The energy momentum tensor is 
\begin{equation}
T_{mn}= \left( \frac{i}{4} e_{(a)  m}\bar{\Psi} \gamma^a D_n \Psi +c.c.\right) + m \leftrightarrow n
\end{equation}
where  the swapping  $ m \leftrightarrow n$  of indexes  is carried for symmetrisation. 
The tetrads of the  Robertson-Walker metric are diagonal:
\begin{equation}
e^{(0)}_0 =1 \quad;\quad e^{(\alpha)}_\beta =a \delta^\alpha_\beta 
\end{equation}
where Greek indexes run over the spatial components and $a=a(t)$ is the cosmological radius scale . In this case the only non-vanishing components of spin-connection are
\begin{equation}
C_{\alpha 0 \beta}=-\dot{a} \delta_{\alpha \beta}
\end{equation}
after some algebra the Dirac Equation reads
\begin{equation}
\left[ i\left( \frac{\partial}{\partial t} -a^{-1}\gamma^0 \vec{\gamma} \cdot  \vec{\nabla} +\frac{3\dot{a}}{2a}\right)-\gamma^0 m\right]\Psi=0
\end{equation}
where $\vec{\nabla}_\alpha =   \partial/\partial x^\alpha$

The generic solution is of the form $\Psi= \Phi(t) e^{-i \vec{k}\cdot \vec{x}}$. For a massless and zero momentum configuration,  $\Psi(t)=\Psi_0 a^{-3/2}$ with $\Psi_0$ a constant spinor. 
The energy-momentum components are
\begin{eqnarray}
T_{00}&=&i \frac{3 H}{4}\Psi^\dagger \Psi +c.c =0\\
T_{\alpha \beta}&=&-i \frac{\dot{a}}{4} \Psi^\dagger \Psi \delta_{\alpha\beta} +c.c=0
\end{eqnarray}
since $ \Psi^\dagger \Psi$ is real . Thus the zero mode (substrate) does not modify the cosmological dynamics.  

Consider now the perturbations generated by the local gravitational field. The  departure of the spinor  from the cosmological background  is here  defined as $\Psi+\delta \Psi =a^{-3/2}(\Psi_0 + \Theta) $ and the tetrad variation $\delta e^{(a)}_m=\varepsilon^{(a)}_m$ such that 
\begin{equation}
\label{hee}
\tilde{h}_{mn}=\varepsilon^{(a)}_m e_{(a) n}+\varepsilon^{(a)}_n e_{(a) n}
\end{equation}

Last, we define $\sigma_{m a b}=\delta C_{m a b}$. One shows that
\begin{widetext}
\begin{equation}
\sigma_{mab}=\frac{1}{2}\varepsilon^{( c) n} \left( e_{(a) n} C_{mab} - e_{(b) n} C_{mac}\right) +\frac{1}{2} \left(e_{(a)}^n \varepsilon_{(b) n;m}-e_{(b)}^n \varepsilon_{(a) n;m}\right) +\frac{1}{2}e_{(b)}^le_{(a)}^p\left(\tilde{h}_{mp;l}-\tilde{h}_{ml:p}\right)
\label{sigma}
\end{equation}

Then the perturbed Dirac equation reads
\begin{equation}
\left[ \left( \frac{\partial}{\partial t} -a^{-1}\gamma^0 \vec{\gamma} \cdot  \vec{\nabla} +\frac{3\dot{a}}{2a}\right)+i \gamma^0 m\right]a^{-3/2}\Theta
=- \gamma^0\left[ \gamma^a   \varepsilon_{(a)}^m\partial_m +\frac{1}{4}\left(\varepsilon_{(a)}^m\partial_m  C_{mbc}  - e_{(a)}^m \sigma_{mbc} \right )\gamma^b \gamma^c \right] \Psi
\label{dirac}
\end{equation}
\end{widetext}

To proceed further we specify the pertubation of the tetrad:
\begin{eqnarray}
2 \varepsilon_{(0)0}&= &\psi \nonumber \\
\varepsilon_{(0)\alpha}+\varepsilon_{(\alpha)0}&=&W_\alpha \nonumber \\
\varepsilon_{(\beta)\alpha}+\varepsilon_{(\alpha)\beta}&=&a f_{\alpha\beta}
\end{eqnarray}
Since  the tetrad $\varepsilon_{(0)a}$ is  time-like, through  a Lorentz transformation we can eliminate all the spatial components $\varepsilon_{(0)\alpha}$. Thus, in this particular Lorentz frame  
$\varepsilon_{(0)\alpha}=0$ and  
\begin{equation}
\varepsilon_{(0)0}=\frac{1}{2} \psi \quad ; \quad \varepsilon_{(\alpha)0}=W_\alpha \quad;\quad \varepsilon_{(\alpha)\beta}=\frac{a}{2} f_{\alpha\beta}
\label{tetrad}
\end{equation}
Inserting these tetrads into eq.(\ref{sigma}), yields
\begin{eqnarray}
\sigma_{00\alpha}&=&\frac{1}{2a} \psi_{,  \alpha}- H W_\alpha 
\label{allsigma} \\
\sigma_{0 \alpha \beta}&=&\frac{1}{2a} \left( W_{\alpha,\beta}-W_{\beta,\alpha}\right) \nonumber \\
\sigma_{\alpha 0 \beta}&=&\frac{a}{2} \left(a^{-1} \left(W_{\alpha,\beta}+W_{\beta,\alpha}\right) -H f_{\alpha\beta}-\dot{f}_{\alpha\beta}-H\psi \delta_{\alpha\beta} \right) \nonumber \\
\sigma_{\alpha \beta \gamma}&=& \frac{a}{2} \left(a^{-1}\left(f_{\alpha \beta,\gamma}-f_{\alpha \gamma,\beta}\right) +H \left (\delta_{\alpha \beta} W_\gamma-\delta_{\alpha\gamma}  W_\beta\right)\right) \nonumber
\end{eqnarray}

Inserting eqs.  (\ref{tetrad}) and (\ref{allsigma}) into (\ref{dirac}), while recalling the substitution $a^{-1} \partial/\partial x^\alpha =\partial/\partial r^\alpha $ leads after some algebra to
\begin{widetext}
\begin{equation}
\left[  \frac{\partial}{\partial t}-\gamma^0 \vec{\gamma}\cdot \vec{\partial}  \right] \Theta 
=\left[\left( \frac{1}{4}\dot{f}-\frac{1}{2} \partial_\alpha W_\alpha\right) 
+\left( \frac{H}{2}W_\alpha +\frac{1}{4} \partial_\beta f_{\alpha\beta}\right)\gamma^0\gamma^\alpha 
 -\frac{i}{4} \partial_\beta W_\alpha \sigma^{\alpha \beta}\right] \Psi_0
\label{intermediate}
\end{equation}
\end{widetext}
where $\sigma^{\alpha \beta}=i[\gamma^\alpha,\gamma^\beta]/2$ and $m=0$. 

The time-dependent solution $\Theta=\theta(\vec{r}) e^{-iEt}$  is not consistent with the rhs, unless $E=0$. This is agreement with the fact that  we regard $\Theta$ as a distortion of the minimum energy configuration $\Psi$ (the substrate) due to the local gravitational field.  Recalling that $\partial/\partial t$ is a derivative with $\vec{x}$-constant of a function that depends on $\vec{r}$,  we can replace $\partial/\partial t\rightarrow H \vec{r}\cdot \vec{\partial}$
\begin{widetext}
\begin{equation}
\left[  H\vec{r}\cdot \vec{\partial}-\gamma^0 \vec{\gamma}\cdot \vec{\partial}  \right] \Theta 
=\left[\left( \frac{H}{4}\vec{r}\cdot \vec{\partial} f -\frac{1}{2} \partial_\alpha W_\alpha\right)  
+\left( \frac{H}{2}W_\alpha +\frac{1}{4} \partial_\beta f_{\alpha\beta}\right)\gamma^0\gamma^\alpha 
\label{spinoreq}
 -\frac{i}{4} \partial_\beta W_\alpha \sigma^{\alpha \beta}\right] \Psi_0
\end{equation}
\end{widetext}

In the spirit of the previous discussions, we solve the equation perturbatively :
\begin{equation}
\Theta=\Theta^{(0)}+ H \Theta^{(1)}+H^2 \Theta^{(1)}+ \cdots
\end{equation}
As in the previous section,  $W_\alpha$ starts at the order$ \sim \mathcal{O}(H)$  (it is related to $T_{0\alpha} $ equation and it vanishes for a static configuration). Then to the lowest order in $H$
\begin{equation}
 \vec{\gamma}\cdot \vec{\partial} \Theta^{(0)} =-\frac{1}{4}  \gamma^\beta \partial_\alpha f^{(0)}_{\alpha\beta} \Psi_0
 \end{equation}
Applying  $\vec{\gamma}\cdot \vec{\partial}$ on both sides
\begin{equation}
\partial^2 \Theta^{(0)} =-\frac{1}{4} \gamma^\mu \partial_\mu \gamma^\beta \partial_\alpha f^{(0)}_{\alpha\beta} \Psi_0
 \end{equation} 
 whose solution is
 \begin{equation}
\Theta^{(0)} = -\gamma^\alpha \gamma^\beta F_{\alpha\beta} \Psi_0
\label{theta}
 \end{equation} 
 where
 \begin{equation}
F_{\alpha\beta}= \frac{1}{16\pi}\int   \frac{\partial'_\alpha  \partial'_\mu {f^{(0)}}_{\mu\beta}' }{\left| \vec{r}-\vec{r}'\right|}d^3r'
 \end{equation}
 and primed functions means they are expressed in terms of $\vec{r}''$. Expanding he spinor equation (\ref{spinoreq}) to the first order in $H$ reads
\begin{widetext}
\begin{equation}
 \vec{\gamma}\cdot \vec{\partial}  \Theta^{(1)} 
=\left[ \vec{r}\cdot \vec{\partial}\gamma^\alpha \gamma^\beta F_{\alpha\beta} +
  \frac{1}{4}\vec{r}\cdot \vec{\partial} f^{(0)} -\frac{1}{2} \partial_\alpha W^{(1)}_\alpha  
 +\frac{1}{4}\partial_\beta f^{(1)}_{\alpha\beta}\gamma^\alpha 
 -\frac{i}{4} \partial_\beta W^{(1)}_\alpha \sigma^{\alpha \beta} \right] \gamma^0\Psi_0
\end{equation}

The energy momentum tensor corresponding to disturbance of the cosmological substrate is

\begin{eqnarray}
\delta T_{mn}&=&\left\{\left[\frac{i}{4} \varepsilon_{(a) m} \left(  \bar{\Psi} \gamma^a \partial_n \Psi+\frac{1}{4} C_{n a b}\bar{\Psi} \gamma^a \gamma^b \gamma^c\Psi\right)+
\frac{i}{16} e_{(a)m}\sigma_{n b c} \bar{\Psi} \gamma^a  \gamma^b\gamma^c \Psi \right. \right. \nonumber \\
&+&\left. \left. a^{-3/2}\frac{i}{4}e_{(a)m} \left( \bar{\Theta} \gamma^a D_n \Psi +\bar{\Psi} \gamma^a D_n \Theta\right) \right] +c. c \right\} \mbox{+m$\leftrightarrow$ n} 
\end{eqnarray}

We are mainly interested in the $\delta T_{00}$ component. Recalling that $C_{0 a b}=0$ , $\Psi(t)=(a_0/a)^{3/2}\Psi_0 $   we get
\begin{equation}
\delta T_{00}=\left(\frac{a_0}{a}\right)^{3/2}  \left[-i\frac{3H}{4} \varepsilon_{(a) 0}  \bar{\Psi}_0 \gamma^a  \Psi_0 
-  \frac{i}{8} \sigma_{0 \beta \gamma} \bar{\Psi}_0 \gamma^0  \gamma^\beta \gamma^\gamma  \Psi_0+\frac{i}{4} \sigma_{0 0 \alpha} \bar{\Psi}_0 \gamma^\alpha \Psi_0+\frac{i}{2} \left(-\frac{3H}{2} \Theta^\dagger  \Psi_0 +{\Psi_0}^\dagger  \partial_0 \Theta \right) \right] +c. c
\end{equation}
\end{widetext}
Now $\bar{\Psi} \gamma^a  \Psi$ is real and  the current $\bar{\Psi}_0 \gamma^\alpha \Psi_0=0$  since there is no preferred cosmological direction.  Furthermore, for a spherical symmetrical configuration $\sigma_{0 \alpha \beta}=0$ [ see eq.(\ref{allsigma})], thus
\begin{equation}
\delta T_{00}=i\frac{H}{2} \left(\frac{a_0}{a}\right)^{3/2}  \left(-\frac{3}{2} \Theta^\dagger  \Psi_0 +\vec{r}\cdot \vec{\nabla}{\Psi_0}^\dagger   \Theta\right) +c. c 
\end{equation}
To the first order in $H$ we need only $\Theta_0$ [eq.(\ref{theta}], 
\begin{equation}
\delta T_{00}  \sim i H F_{\alpha \beta} \Psi_0^\dagger \gamma^\alpha \gamma^\beta \Psi_0 +cc 
\end{equation}
Clearly, in a spherical symmetrical configuration $F_{\alpha \beta}$ is symmetric, thus
\begin{equation}
\delta T_{00}  \sim i H F \Psi_0^\dagger \Psi_0 +c.c =0 \quad ,
\end{equation}
where $F=\sum_\alpha F_{\alpha \alpha}$. Accordingly, a spinor cannot induce a first order in $H$   correction to the Newtonian potential.

Unforseenably, none of  the field  theories  studied in this paper can produce odd  corrections in $H$ to the local gravitational fields and therefore, cannot bring about substantial corrections to the local dynamics.

In the lack of a general   principle forbidding  odd powers in $H$, it is conceivable that some field theory could bring about odd powers in the $H$-expansion.  Should such a theory exist,  the lowest order corrections are linear in $H$  and  on dimensional grounds
\begin{equation}
\psi \sim - \frac{MG}{r} + H r +H MG \ln (r ) + \dots
\end{equation}
Accordingly, the velocity profile, away from the mass distribution would be
\begin{equation}
v^2 \sim \frac{ M G }{r} + Hr +  MG H +\cdots
\end{equation}

The last term yields flat rotation curves, but  comparing to the Newtonian term reveals that it becomes relevant only at scales $ r_0 \sim H^{-1}$, thus meaningless. The  second term gives a linearly growing velocity curve at  a very much small slope such that could be mistakenly taken for a flat rotation curve at galactic scales.  Furthermore, comparison with the Newtonian potential reveals that it becomes relevant at scales $r_0 \sim (M G/H)^{0.5} \sim 5 kpc $ for a typical galaxy.  At the $r  \sim r_0$ region where there is  dynamical transition from the Newtonian behaviour to the $H r$ term the velocity scales is $ v_0^4 \sim  M^2 G^2 /r_0^2 \sim MG H$, which is  nothing but Tully-Fisher's Law ! Furthermore,  the corresponding  acceleration scale in this region $a_0 \sim v_0^2/r_0 \sim H$.  Needless to say the utmost importance of   scrutinizing   field theories that could bring about  linear corrections in $H$ to the gravitational potential  or either showing  that   odd term corrections are forbidden.

\subsection*{acknowledgements}
I am grateful to Prof. J. D. Bekenstein for enlightening discussions.

\end{document}